\documentclass[11pt,twoside]{article}
\usepackage{newpasp}

\usepackage{epsf}

\index{Mihos, J.C.}

\begin{document}

\title{Do mergers make (normal) ellipticals?}
\author{Chris Mihos}
\affil{Case Western Reserve University, Cleveland, OH 44106}

% A concise abstract is recommended.  Enter the text of the abstract in
% between the \begin{abstract} and \end{abstract} commands.  Do NOT
% include the word ``Abstract'' in your text; it is insterted
% automatically. Do NOT  make a paragraph break between \begin{abstract} 
% and the first line of the text of the abstract!  Abstracts are required 
% for all papers.

\begin{abstract}

Under the merger hypothesis, elliptical galaxies are built through
mergers of gas-rich spirals. However, the relative paucity of HI in
most normal ellipticals demands significant processing of atomic gas
into other forms if this process is to be viable. Here I present a few
qualitative remarks on how the properties of merger-spawned
ellipticals might depend on the (evolving) gas content of the
progenitor disks, then turn to a more quantitative study of the
constraints provided by the nuclear properties of ellipticals and
merger remnants.

\end{abstract}

% Include keywords if you wish. The keywords.apj file, found on aas.org 
% in the pubs/aastex-misc directory, contains a list of keywords used 
% with the ApJ and Letters.  

%\keywords{meetings: good ones -- conferences: gas -- conferences: galaxies}

% That's it for the front matter.  On to the main body of the paper.

%The realization that the remnants of disk galaxy mergers possess many
%properties in common with elliptical galaxies has lead to the ``merger
%hypothesis'' which states that the present-day population of
%ellipticals was formed via mergers of disk galaxies at higher
%redshift. Testing the merger hypothesis has typically involved
%comparing nearby merger remnants to ``normal'' ellipticals, looking
%for discrepant properties that might distinguish merger-spawned
%ellipticals from those built by other processes (e.g., monolithic
%collapse models). However, in a universe which grows hierarchically,
%{\it all} structure grows through accretion; indeed, instead of acting
%as any absolute test of the merger hypothesis, differences between
%nearby mergers and ellipticals may tell us more about the detailed
%accretion history of galaxies. In this light, the salient question may
%not be ``do mergers make ellipticals?'' but rather ``what merged when
%to make ellipticals?''

\section{Properties of Merger Remnants - Evolution or Prevolution?}

Tests of the merger hypothesis have typically taken the form of a
comparison between the properties of merger remnants and normal
ellipticals. Such studies have pointed towards two possible
discrepancies related to the hydrodynamic evolution (and subsequent
aftereffects) of merging galaxies. First, X-ray observations have
shown that the hot gas halos surrounding young merger remnants are
quite modest -- the $L_x/L_B$ ratios for remnants are more typical of
those of spiral galaxies than of ellipticals (Read \& Ponman 1998;
Sansom et al 2000). Second, recent studies of globular cluster
populations in mergers hint that young merger-spawned ellipticals may
be deficient in globular clusters when compared to elliptical galaxies
(Grillmair et al 1999; Brown et al 2000).  Subsequent evolution of the
remnants will bring these properties more in line with normal
ellipticals: the fading of stellar populations will increase the
globular cluster specific frequency, while mass loss from evolving
stars may subsequently form a hot gaseous halo. However, the evolution
must be significant and rapid to make these objects look like normal
ellipticals in less than a Hubble time.

Here I argue that in addition to evolution, adding a bit of
``prevolution'' to the scenario may help explain these apparent
discrepancies.  Much of our understanding of the detailed evolution of
merging galaxies comes from low redshift. Even dynamical simulations
have largely focussed on models with merging progenitors similar to
those of nearby disk galaxies.  When we compare the models to nearby
mergers such as the Antennae or NGC 7252, the comparison is
well-founded. However, if normal ellipticals were made from mergers at
higher redshift, the progenitors may have well been different from
galaxies in the local universe. One of the simplest expectations is
that mergers at higher redshift may have involved galaxies with a high
gas fraction. As an example, in an $\Omega_M=0.3, \Omega_\Lambda=0.7$
cosmology, a spiral galaxy formed at $z_f=3$ with an exponentially
decaying star formation rate with decay timescale $\tau=5$ Gyr has a
gas fraction of $f_g=0.1$ at $z=0$ and $f_g=0.5$ at $z=1$. The fact
that the gas fraction of galaxies is changing with time may lead to
systematic differences between the properties of merger-spawned
ellipticals of different ages.

Ideally, we would be able to model the complete evolution of the
gaseous phase of galaxies to ask how hot halos and young globular
clusters form. Such a task has proved exceptionally difficult, due to
a combination of physical and computational
limitations. Qualitatively, however, it seems that the evolving gas
fraction might explain both the low $S_N$ and low $L_x/L_b$ of
morphologically young ellipticals. In both cases, if the ``processing
efficiency'' (i.e., the fraction of gas processed into globular
clusters or into hot gas through starburst winds or shock heating) is
fixed, a falling gas fraction would necessarily result in the
formation of low $S_N$, low $L_x/L_b$ over the course of cosmic
history. Of course, subsequent evolution likely {\it will} move these
quantities back towards those of ``normal'' (i.e., older) ellipticals,
but there is no need to demand that they {\it match} those older
ellipticals. In this sense, rather than being an indictment of the
merger hypothesis, discrepancies between the properties of young
merger-spawned ellipticals and older normal ellipticals may actually
tell us about the effects of galaxy ``prevolution'' on the merging
process.

% We could have also have used the \plotfiddle command
% \plotfiddle{file}{vsize}{rot}{horiz scale}{vert scale}{dx}{dy}
 
%% TO INCLUDE EQUATIONS OR TABLES, PLEASE SEE THE FILE
%% newpaspman.ps at http://www.aspsky.org/pubs/authors.html

\section{The Central Parameter Relationship for Merger Remnants}

Theoretical arguments indicate that it is in the nuclei of remnants
where the merger hypothesis may face its most stringent test, as a
wide variety of physical processes may act to shape their
nuclei. Dynamical heating, gas dissipation and star formation, black
holes (and black hole binaries), and low mass accretion can all act to
alter the central densities of galaxies in different ways. Yet studies
of elliptical galaxy nuclei reveal very well-behaved relationships
between their nuclear and global properties.  In particular, the
so-called ``central parameter relationship'' shows that elliptical
galaxies show a strong anticorrelation between their central densities
and their total luminosities ((Lauer et al 1995; Faber et al 1997
(F97)).  It is a non-trivial question whether or not such a
correlation is consistent with a scenario where mergers drive the
formation of elliptical galaxies. For example, two of the most
discrepant points on the central parameter relationship of F97 are NGC
1316, the central galaxy in the Fornax cluster and a clear merger
remnant, and NGC 4486B, a close companion to M87 possessing a double
nucleus.  If merger remnants in general show marked deviations from
the central parameter relationships, the status of the merger
hypothesis as a mechanism for forming the majority of normal
elliptical galaxies would be in serious doubt.  As such, the
robustness of the central parameter relationship may provide a strong
test of the merger hypothesis.

We (van der Marel et al, in preparation) have used NICMOS to investigate the
central parameter relationship in a sample of young ellipticals with
significant tidal debris. Our sample is chosen morphologically from
the Arp (1966), Vorontsov-Velyaminov (1977), and UV-bright Markarian
(Mazzarella \& Boroson 1993) catalogs. Our sample is chosen to have
galaxies with tidal features indicative of a recent merger -- tails,
shells, plumes, or otherwise strongly distorted isophotes, and
includes the well-known merger remnants NGC 7252 and NGC 3921. The
sample is also restricted to  $z<10,000$ km/s, which
translates to a spatial resolution of ${_<\atop^{\sim}}$ 25 pc (for
$H_0=80$) using the NIC1 camera.

\begin{figure}
\plotfiddle{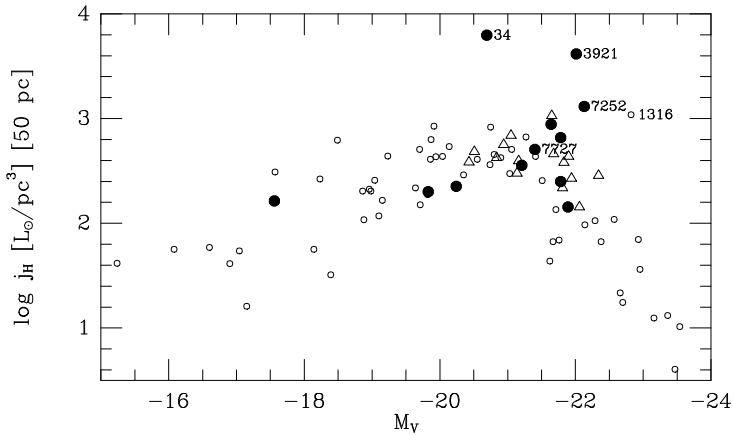}{2.0in}{0.}{50}{50}{-150}{-210}
\caption{\small The central parameter relationship for normal
elliptical galaxies (open circles; from Faber et al 1997),
counter-rotating core galaxies (open triangles; from Carollo et al
1997), and our sample of morphologically peculiar ellipticals (filled
circles; van der Marel et al, in preparation).  The central luminosity
densities for normal and counter-rotating core ellipticals have been
converted to the infrared assuming $V-H=3$, typical for elliptical
galaxies (Peletier et al 1990; Silva \& Bothun 1998).}
\label{fig1}
\end{figure}

Each system was imaged using the NIC2 camera using the F110W, F160W,
and F205W filters, and again in F110W using the NIC1 camera for
maximum resolution. The images were Lucy deconvolved using the
appropriate PSF, after which azimuthal surface brightness profiles
were extracted.  The surface brightness profiles were then fit by a
``nuker law'' (Lauer et al 1995) and deprojected to obtain the three
dimensional luminosity density. Figure 1 shows the H-band luminosity
density at $r=50$ pc for our sample, plotted as a function of galaxy
luminosity.  The majority of the objects in our sample lie on the same
density-luminosity relationship defined by normal ellipticals. Three
of the objects, however, show significantly higher central luminosity
densities than would be expected from the normal elliptical
density-luminosity relationship: NGC 34, NGC 3921, and NGC 7252. The
scatter in central luminosity density in our sample is significantly
larger than that seen in normal elliptical samples.

This large scatter is likely the result of a number of effects. First,
our sample is morphologically diverse, comprised of galaxies which
have suffered a variety of interactions. It is interesting that the
three galaxies which show the central luminosity excess have the most
prominent tidal debris, suggesting that {\it major mergers} are the
most likely to affect the central luminosity density of
galaxies. However, there is no one-to-one correlation between tidal
morphology and nuclear properties: a fourth object in our sample, NGC
7727, also possess a long, prominent tidal tail, yet shows no excess
nuclear light. A second effect also likely plays a major role: that of
merger age. Their blue nuclear colors suggests that the high
luminosity densities observed in NGC~34, 3921 and 7252 are probably a
direct consequence of recent star formation triggered by a
merger. Stellar populations fade with time, and these galaxies will
therefore evolve towards the locus of normal elliptical galaxies as
time passes.  Tidal debris becomes less prominent with age as well, as
the tidal tails expand away or fall back and mix into the remnant
(Mihos 1995; Hibbard \& Mihos 1995). Unfortunately, it is very
difficult to disentangle the effects age and encounter type; {\it
both} effects are almost certainly at work in our morphologically
selected sample.

At face value, the large scatter in the nuclear properties of our
sample of merge remnants might seem a blow against the merger
hypothesis for the formation of ellipticals. However, the
morphological selection criteria biases us towards specific types of
mergers: major, prograde encounters, which are the most effective at
triggering strong nuclear inflows. Mergers in general will sample a
wider range of encounter parameters which may not be as efficient
at altering the nuclear properties. Furthermore, as noted above,
evolution of the stellar populations will likely drive the discrepant
objects to lower luminosity and central luminosity density, moving
them back towards the mean relationship with time. But this evolution
is slow; it will take several Gyr for the starburst population to fade
sufficiently. An object like NGC 7727, whose prominent tidal debris
argues for a young age, cannot have started with a central luminosity
spike as strong as that observed in NGC 3921 and have evolved so
quickly onto the central parameter relationship.

Peculiar ellipticals with strong central density spikes may be the
natural evolutionary outcome of the ultraluminous IRAS galaxies. These
luminous merger induced starbursts possess very strong central
concentrations of gas, and if that gas is converted efficiently into
stars the resulting central density will be quite high (Hibbard \& Yun
1999).  Because it can take a Hubble time to evolve back onto the
central parameter relationship, our results suggest either that most
ellipticals do not go through this ultraluminous central starburst
phase, or that the merger ages of most ellipticals must be very
large. Our results are inconsistent with the notion that ULIRGs at
moderate redshifts formed a significant fraction of the local elliptical galaxy
population.

\end{document}